\documentclass[article]{revtex4}

\pdfoutput=1
\usepackage[pdftex,
]{hyperref}
\medskip

\begin{document}

\begin{flushright}
{
SLAC--PUB--12507\\
May, 2007\\}
\end{flushright}

\author{Travis C. Brooks}
\affiliation{Stanford Linear Accelerator Center Library/SPIRES
  Databases\\Stanford University\\Stanford, CA 94309}
\title{Open Access Publishing in Particle Physics: \\ A Brief Introduction
  for the non-Expert \footnote{Work supported by Department of Energy contract
DE--AC02--76SF00515}}

\begin{abstract}

Open Access to particle physics literature does not sound particularly new
or exciting, since particle physicists have been reading preprints for
decades, and arXiv.org for 15 years.  However new movements in Europe are
attempting to make the peer-reviewed literature of the field fully Open Access.  This is
not a new movement, nor is it restricted to this field.  
However, given
the field's history of preprints and eprints, it is well suited to a
change to a fully Open Access publishing model. Data shows that 90\% of
HEP published literature is freely available online, meaning that HEP
libraries have little need for expensive journal subscriptions.  As libraries begin to cancel journal
subscriptions, the peer review process will lose its primary source of
funding. Open Access publishing models can potentially address this issue.  European
physicists and funding agencies are proposing a consortium, SCOAP3, that
might solve many of the objections to traditional Open Access publishing
models in Particle Physics.  These proposed changes should be viewed as a
starting point for a serious look at the field's publication model, and
are at least worthy of attention, if not adoption.

\end{abstract}

\maketitle

In November 2006 a meeting of European particle physics laboratories,
funding agencies, librarians, and researchers took place at CERN.
After the meeting it was announced \cite{CERNPR} that an interim working party had been
formed to proceed with an initiative known as the
Sponsoring Consortium for Open Access Publishing in Particle Physics
(SCOAP3).  This announcement was the latest of several that
have begun to make the particle physics community more aware of the
the Open Access movement and its relation to the publication of
physics literature.  However there are still many who are
unaware of the issues involved, and of the new directions suggested by
the European particle physics community.   An overview of 
this movement and its relation to particle physics is presented herein.
It is necessarily an oversimplification of some issues, and the reader is urged to consult
the references herein, or your local librarian to find out more
detail.    

The Open Access (OA) movement has been formalized over the past several
years to advocate for a change in the methods
involved in the distribution of academic literature \cite{OATIME}. 
In short, an Open Access article can be roughly defined
as one that is freely available to read immediately after traditional 
peer-review, in
perpetuity, and with unrestricted use\cite{OADEF}.  This is in contrast to most
scientific literature, which currently requires a subscription fee in
order to read the article online or receive a hard copy by mail.

Note that Open Access is not the same as electronic access.  Many
researchers are able to read all relevant articles online as soon as
they are published, but this is due to libraries paying fees for online
(and print) subscriptions.  In recent years these
subscription costs have been rising precipitously \cite{SERIAL}, causing 
some to
worry about the sustainability of the communication model wherein authors
generate papers, and libraries pay to obtain access to them.  As library
budgets are pressured and serials costs grow, some journals are
necessarily cut and researchers lose access to literature. This area of
concern is known as the serials crisis.  Open Access advocates often speak
of resolving the serials crisis \cite{OASER}, though that is by no means the sole,
or even primary, goal of the movement.  

Harnad et.al. describe two roads to open access, the green and the
gold. \cite{GREGOL}  The gold road is the formation and support of Open Access
journals.  This is the focus of the recent discussion in Europe, and is 
the most relevant for particle physics. However, as background, one should
understand the so-called {\it green road}, also known as {\it self-archiving}. This is simply an author
posting preprint (and post-print) copies of their papers online on her
own repositories, on eprint archives such as arXiv.org, or in
institutional repositories \cite{GREENR}. Authors can achieve the laudable goal of
providing universal, free access to their work without worrying about the
business models of publishers. Authors should be excited about this
prospect because their work is then more visible to other researchers, and
in turn they can more easily find and read works of interest to them.  Note
that this does not solve the serials crisis at all, but perhaps once this level
of access has been achieved, journal business models may change in
response.

In Particle Physics, self-archiving  has existed for 15 years using
arXiv.org as a repository for articles, and SPIRES as the search engine
that unifies the literature\cite{SPIRES}.  By
studying SPIRES data one finds that the fraction of published articles on arXiv
is well over 95\% for large mainstream journals like Nucl. Phys. B and
Phys. Rev D.  Smaller, less mainstream journals tend to have a
lower arXiv rate, but in general around 90\% of published, peer-reviewed HEP
literature (not including conference proceedings) from the last 10 years is available online
freely at arXiv.  
Thus particle
physics, unlike other fields, can essentially ignore calls to self-archive
its papers, since this has already been accomplished.

However, HEP libraries still face a serials crisis. One might propose that
a simple solution would be to cancel all journals that are solely
particle physics, since the content is essentially freely available. However,
this is not occurring, and more to the point, this would be a disaster
for the field if it did occur. Physicists continue to publish as much
as before self-archiving was a reality.  In fact in the hep-th section of
arXiv.org, around 75\% of the articles are eventually published in
journals \cite{SPIDAT}.  Why do they publish? It isn't certain, but one
can see that journals are not used for communication within the field, due
to the high rate of arXiv submissions. 

Because journals are not used for communication, one of their only
remaining uses is to provide
peer-review.  If particle physics journal subscriptions were all
canceled, the mechanism for peer-review would be jeopardized.  So, while
particle physics clearly already has Open Access in the ``green road'' sense,
the field has an odd model where libraries essentially subsidize the cost of
peer-review and other functions of journals.  Since physicists can read
articles without these subscriptions, this subsidy is entirely voluntary
on the part of the libraries, but is necessary for the field as a whole.
As libraries face increasing journal costs, they have little incentive to
continue to buy journals that are not used for communication.  Libraries
currently fund essentially 100\% of the peer-review process in HEP\cite{VOLREF}, while
the product that they get in return for their payment is 90\% freely
available on arXiv.org. As
university and laboratory libraries begin to look more closely at their
serials collections, these subsidies for the peer-review process may begin
to disappear.   

Enter the ``gold road'' or Open Access journal.  There are several different
business models for a an Open Access journal, but it cannot, by
definition, use the ``pay to read'' model. The ``pay to read'' model made
sense when the cost of distribution scaled with the number of copies produced.
Now, since most dissemination can be handled electronically, the
primary costs (peer-review, servers, etc) scale with the number of
articles published or submitted.  Thus some journals are shifting 
to an ``pay for publication'' model of dissemination in which authors of
research articles (more often their funding agencies) pay a fee to a
publisher who in turn reviews the article and disseminates it to the
public for free.  This puts the burden of paying for peer-review back on
the authors and their funding agencies, i.e. the people who generate the
research that requires, and benefits from, review.   Other models of Open
Access journals include direct funding from grants, advertising, and
charging for related/add-on products.  For example, Phys. Rev. Special
Topics: Accelerators and Beams is Open Access via a
sponsorship model, whereby accelerator laboratories worldwide sponsor the
journal so no subscription is needed\cite{PRSTAB}.  However, the ``pay to publish'' or
``author pays'' model is the most common, and the one of the most relevant
to particle physics.   

There are arguments against the ``pay to publish'' implementation of
Open Access in journals (see, e.g. \cite{GREENR},\cite{NTLACD}). These include:
\begin{itemize}
\item{Conflict of interest: if a journal gets money from its authors,
  might it not treat the peer-review process differently? Quality
  might diminish.  Yet it is not clear what a journal would gain in the
  long term by publishing low-quality material.  Impact factors and
  community judgment would quickly remove any temporary financial incentive to do this.}

\item{Poorer researchers/institutions: Poorer
  researchers might not be able to pay open access author fees.
  It should also be noted that most author-pays models propose to operate
  at the level of   line-items on grants, so that the relative wealth of
  an individual   author or institution is irrelevant in practice.  At a
  national level,   there are current programs that work to reduce the
  costs of journals in   developing nations\cite{DEVEL}, and these practices could
  carry over into a pay to publish model.   Further, it should be noted that
  a pay to publish model shifts the financial burden of reviewing the
  research literature to the   producers of that literature. This implies
  that larger institutions, which produce large amounts of research,
  will support a larger fraction of peer-review costs for the field.  This
  seems as though it would benefit, rather than hurt, small
  institutions. However, smaller institutions still may have very active
  theory groups which produce 
  a large amount of research output.  This might lead to a problem under a
  simple author-pays model, since theorists write the vast majority of
  the papers in the field \cite{EXPTHE}, but aren't necessarily at major institutions.
  Finally note that Open Access to the literature itself 
  (via either road) does nothing but help poorer institutions, by
  removing the cost of obtaining scientific literature from their research
  overheads.  }

\item{Authors won't pay: if scientists are forced to choose 
  between paying for research needs (graduate students, equipment,
  etc.) and strange new open access fees, almost all will choose
  research.  Hence any pay to publish model must provide institutional 
  and funding agency policies that not only recommend or require open access
  publication, but also provide funds earmarked for this purpose.  It
  may even be preferable to use libraries and/or some other external
  infrastructure to pay these costs, so that authors need not worry
  about new details.}

\item{Libraries pay twice: If libraries continue their subscriptions, which
of course they must until a large quantity of literature is open access,
then they are paying the publisher for articles that the author already
paid for access to.  Only if journal subscriptions are canceled or prices
drop with the fraction of open access articles will there be a cost
savings.  In any case there is a transitional period during which there is
probably no way to avoid some extra costs.  If a journal completely transitions to
an Open Access model, then this objection is
eliminated, since there is no longer any subscription fee. However, if the
journal publishes some articles of each type without decreasing
subscription costs, or if it bundles OA journals with non-OA journals in
package deals, then this double payment
could occur. }

\end{itemize}

It seems clear, though not unambiguously so,  that open access is a
general good for the research community.  Further it seems that, in a
field which has Open Access to its literature,  a ``pay to publish''
model for peer-review costs makes more sense than a ``pay to read'' model.    
Particle physics is thus in
a position where the transition to a ``pay to publish'' model might make some
sense, and might be particularly easy.  

With this background one can now understand the role and significance
of SCOAP3 consortium recently proposed by European funding agencies \cite{SCOAP}, \cite{CERNPR},\cite{CERNOA}. 
 With the start of the LHC at CERN there is an opportunity to transition all of
the literature of the field to the ``pay to publish'' model.  SCOAP3 proposes to
pay author fees for Open Access in a ``pay to publish''  model for a large
segment of the physics literature. The proposal would make the change
transparent to authors themselves as SCOAP3 and the main funding agencies
shift subscription costs to cover publication charges centrally. \cite{MELE}
 SCOAP3 would essentially provide bridge funding to help authors,
journals, and libraries transition to this model, preventing much of the
double charging that might occur.  After a transition period of 3-5
years one would hope that libraries and/or funding agencies would be paying
publication charges, rather than subscriptions, for all particle physics
journals, and libraries would no longer need to voluntarily subsidize the
peer-review services of journals. SCOAP3 might continue to exist after the
transition as well, as an umbrella consortium for libraries and funding
agencies funding Open Access payments.\cite{CERNOA}

Note that this movement is not aimed at specific journals or publishers.
At present almost all relevant journals currently provide authors with
an option to pay for OA (costs from \$900-\$3000/article), or are ready to
implement such a structure immediately.  This includes Phys Rev D, Phys
Rev Lett, JHEP, Nucl Phys B, Phys. Lett. B, Eur. J. Phys, Nucl Inst. Meth,
J Phys G, and others.  There are also fully Open Access journals emerging like New
Jour. Phys. and PhysMathCentral. Authors and funding agencies who wish to can pay for
their articles to be Open Access in most of these journals today. Authors,
of course, should not generally be expected to volunteer to pay this cost, any more
than libraries or publishers, so funding agencies that fund both
subscription costs and authors doing research must be the instruments of change. For real change to happen the movement needs to include not only
Europe but the United States, Asia, and the entire world. To this end
libraries, authors,
and funding agencies should make themselves aware of the journals'
policies, the proposed role of SCOAP3, and the arguments for and against
such a move.  These issues demand attention and action from all parties
involved in the production and dissemination of particle physics
literature.

Regardless of your position on open access and the ``pay to publish'' model,
it is clearly an exciting time in particle physics communication,
and change is almost certainly on the horizon. Finally, thanks are due to
the people who have advocated and organized to help prepare the field to
make this change.  Open Access in particle physics was a reality 15 years ago due to arXiv.org and SPIRES.  Now,
thanks to CERN and others, the field has the unique opportunity to change
its publication model in a way that might match its communication model


\begin{thebibliography}{99}
\bibitem{VOLREF} With the exception of volunteer referees, there is no
  other funding source for most existing peer review.
\end{thebibliography}
\end{document}